%% file: spl_article.tex
\documentclass[journal]{IEEEtran}

\ifCLASSINFOpdf
\else
   \usepackage[dvips]{graphicx}
\fi
\usepackage{url}

\usepackage{graphicx}
\usepackage{amsmath, amssymb, xfrac}
\usepackage{hyperref}

\begin{document}

\title{Improving Numerical Stability of Normalized Mutual Information Estimator on High Dimensions}

\author{Marko Tuononen, \IEEEmembership{Member, IEEE}, and Ville Hautamäki, \IEEEmembership{Member, IEEE}
\thanks{Manuscript submitted for review 9 October 2024. Revised manuscript submitted for review 6 June 2025.}
\thanks{Marko Tuononen is with Nokia Networks, P.O. Box 226, 00045 Nokia Group, Finland, and also with School of Computing, University of Eastern Finland, P.O. Box 111, 80101 Joensuu, Finland (e-mail: marko.1.tuononen@nokia.com).}
\thanks{Ville Hautamäki is with School of Computing, University of Eastern Finland, P.O. Box 111, 80101 Joensuu, Finland (e-mail: ville.hautamaki@uef.fi).}}

\markboth{Preprint version. Submitted to IEEE Signal Processing Letters, July 2025} {Tuononen \MakeLowercase{\textit{et al.}}: Improving Numerical Stability of Normalized Mutual Information Estimator on High Dimensions}

\maketitle

\begin{abstract}
   Mutual information provides a powerful, general-purpose metric for quantifying the amount of shared information between variables. Estimating normalized mutual information using a k-Nearest Neighbor (k-NN) based approach involves the calculation of the scaling-invariant k-NN radius. Calculation of the radius suffers from numerical overflow when the joint dimensionality of the data becomes high, typically in the range of several hundred dimensions. To address this issue, we propose a logarithmic transformation technique that improves the numerical stability of the radius calculation in high-dimensional spaces. By applying the proposed transformation during the calculation of the radius, numerical overflow is avoided, and precision is maintained. Proposed transformation is validated through both theoretical analysis and empirical evaluation, demonstrating its ability to stabilize the calculation without compromising precision, increasing bias, or adding significant computational overhead, while also helping to maintain estimator variance.   
\end{abstract}


\IEEEpeerreviewmaketitle

\section{Introduction}
\label{sec:introduction}

\IEEEPARstart{M}{utual} Information (MI) is a fundamental concept in information theory \cite{kullback1959information, cover2005, murphy2022}, widely used in machine learning \cite{gokcay2002, strehl2003, paninski2003, peng2005}, signal processing \cite{maes1997, hudson2006}, and statistics \cite{veyrat2009, reshef2011, ryan2016}. Unlike correlation-based metrics, MI captures both linear and non-linear dependencies between random variables. Accurately estimating MI from finite samples remains a central challenge \cite{czyz2024pmi,abdelaleem2025}.

MI quantifies the reduction in uncertainty about one variable given knowledge of another variable. It is defined as the Kullback-Leibler divergence between the joint distribution and the product of the marginals \cite{murphy2023}:
\begin{equation}\label{mi_definition}
   \phantom{.}I(X; Y) \triangleq D_{KL} \left(p(x, y) \mid\mid p(x)p(y)\right).
\end{equation}

For discrete random variables, MI is calculated as \cite{cover2005}
\begin{equation}\label{mi_equation_discrete}
   \phantom{,}I(X; Y) = \sum_{x \in X} \sum_{y \in Y} p(x, y) \ln \frac{p(x, y)}{p(x)p(y)},
\end{equation}
or via Shannon entropy \cite{shannon1948}:
\begin{equation}\label{mi_via_entropy_definition}
   \phantom{.}I(X; Y) = H(X) + H(Y) - H(X; Y).
\end{equation}

For continuous variables, summations are replaced with integrals using \textit{differential entropy} \cite{shannon1948}:
\begin{equation}\label{mi_equation_continuous}
   \phantom{.}I(X; Y) = \int_{x \in X} \! \int_{y \in Y} p(x, y) \ln \frac{p(x, y)}{p(x)p(y)} dx \, dy.
\end{equation}

Estimating MI for continuous variables is difficult due to the need to estimate joint densities, especially in high dimensions where the curse of dimensionality \cite{bellman1961} causes severe sparsity. Traditional estimators like histograms \cite{pizer1987, scott1979} and kernel density methods \cite{moon1995} degrade in such regimes. \textit{k-Nearest Neighbor (k-NN) estimators} \cite{kozachenko1987, kraskov2004, nagel2024nmi} are more adaptive, while recent neural and variational estimators \cite{belghazi2018, lin2019, poole2019, song2020} show promise for complex distributions, though challenges with sparsity and long tails remain \cite{czyz2023}.

\begin{figure}
   \centerline{\includegraphics[width=\columnwidth]{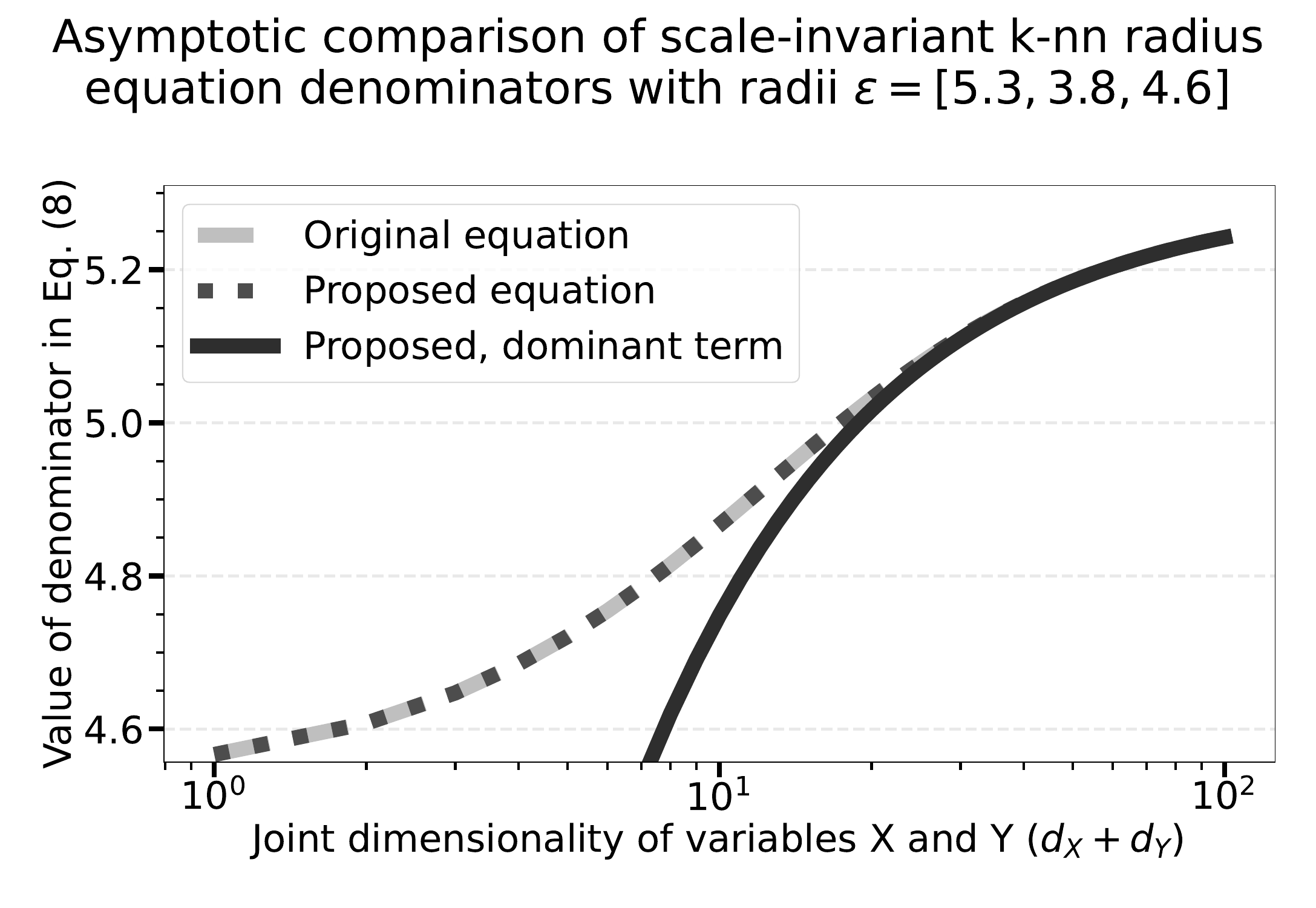}}
   \caption{Comparison of the original, proposed, and dominant-term equations of the scaling-invariant k-NN radius in Equation~\eqref{scaling_invariant_radius_equation}, showing asymptotic behavior of denominator values $V$ as a function of joint dimensionality $d_X + d_Y$ for fixed radii $\varepsilon$.}
   \label{fig:asymptotic_comparison}
\end{figure}

In this work, we adopt the \textit{Kraskov-Stögbauer-Grassberger (KSG) estimator} \cite{kraskov2004}, which uses Equation~\eqref{mi_via_entropy_definition} and entropy values derived from the k-NN algorithm \cite{cover1967}. For normalization, we apply the method of Nagel et al. \cite{nagel2024nmi}, which is invariant to variable transformations and provides an upper bound on MI. However, it becomes numerically unstable in high-dimensional settings, particularly when computing the scaling-invariant k-NN radius in Equation~\eqref{scaling_invariant_radius_equation}.

To address this, we propose a logarithmic transformation that stabilizes this computation without loss of precision, as illustrated in Figure~\ref{fig:asymptotic_comparison}. Introduced as the \texttt{volume\_stable} measure in \cite{tuononen2024volumestable}, our method improves the robustness of MI estimation for high-dimensional, large-scale datasets. This enhances applicability in areas such as computational neuroscience \cite{palmer2015}, bioinformatics \cite{mehtonen2019,demirtas2024}, and multivariate statistics \cite{varley2024}---as well as other fields discussed earlier. Compared to alternatives like approximate computing \cite{han2013approximate,liu2022approximate} or arbitrary-precision arithmetic \cite{knuth1997seminumerical,fousse2007mpfr}, our approach improves numerical stability without adding bias or significant computation, and reduces variance by preventing numerical instability in high-dimensional settings.

\section{Background}
\label{sec:background}
The KSG estimator \cite{kraskov2004} uses k-Nearest Neighbors (k-NN) \cite{cover1967} to estimate joint and marginal entropies directly from the data. For each data point, the algorithm calculates the distance to its k-th nearest neighbor in joint space and counts the number of points within that same distance in the marginal spaces. These distances and counts are used to estimate the densities and entropies of the joint and marginal distributions, making the estimator non-parametric and independent of explicit probability distribution assumptions. The KSG estimator is expressed as:
\begin{equation}\label{ksg_estimator}
\begin{split}
   \widehat{I(X; Y)} = \psi(N) + \psi(k) - \langle \psi(n_X+1) + \psi(n_Y+1) \rangle,
\end{split}
\end{equation}
where $N$ is the number of samples, $k$ is the number of nearest neighbors, $n_X$ and $n_Y$ represent the number of points within the marginal spaces of $X$ and $Y$ that are within the same radius as the distance to the k-th nearest neighbor in the joint space of $(X; Y)$, and $\psi(\cdot)$ is the digamma function. The angle brackets $\langle \cdot \rangle$ indicate an arithmetic mean over all data points. The definition of digamma function can be found from mathematical handbooks, such as \cite[pp. 258--259]{abramowitz1972digamma}.

\subsection{Normalizing Mutual Information}
\label{ssec:normalizingmutualinformation}
Normalized Mutual Information (NMI), as defined in \cite{nagel2024nmi}, is the mutual information $I(X; Y)$ normalized by the geometric mean of the marginal entropies $H(X)$ and $H(Y)$:
\begin{equation}\label{nmi_equation}
   \phantom{.}I_{\text{NMI}}(X; Y) \triangleq \frac{I(X; Y)}{\sqrt{H(X) H(Y)}}.
\end{equation}

However, this formulation applies only to discrete entropies. For continuous variables, differential entropy is not invariant under coordinate transformations, which can lead to negative entropy values. Jaynes \cite{jaynes1968} introduced \textit{relative entropy} to resolve these issues, incorporating an invariant measure $m(x)$ to correct for scale and parametrization changes:
\begin{equation}\label{relative_entropy_definition}
   \phantom{.}H_r(X) = - \int_{x \in X} p(x) \ln  \frac{p(x)}{m(x)} dx.
\end{equation}

Nagel et al. \cite{nagel2024nmi} applied the concept of relative entropy to ensure that NMI and its normalization factor in \eqref{nmi_equation} remain consistent across different coordinate systems by using invariant measures derived from local volumes around data points.

\subsection{Scaling-Invariant k-NN Radius and Entropy Estimation}
\label{ssec:scalinginvariantradiusestimation}
A critical aspect of Nagel et al.’s method \cite{nagel2024nmi} is the calculation of the k-NN radii, $\varepsilon$, which estimates the local densities of data points. To avoid bias from scale or dimensionality, the raw k-NN distances are normalized:
\begin{equation}\label{scaling_invariant_radius_equation}
   \phantom{,}\tilde{\varepsilon} = \frac{\varepsilon}{ \langle \varepsilon^{d_X + d_Y} \rangle^{1/(d_X + d_Y)} } = \frac{\varepsilon}{V},
 \end{equation}
where $V = \langle \varepsilon^{d_X + d_Y} \rangle^{1/(d_X + d_Y)}$ is the normalization factor, and $d_X$ and $d_Y$ are the dimensionalities of the marginal spaces.

\subsection{Entropy Estimators with Invariant Measures}
\label{ssec:entropyestimators}
Nagel et al. \cite{nagel2024nmi} developed entropy estimators that incorporate the scaling-invariant k-NN radii from Equation~\eqref{scaling_invariant_radius_equation}. These estimators use relative entropy to ensure that marginal and joint entropy estimates are invariant across different scales and transformations. The marginal and joint relative entropy estimators are given as:
\begin{equation}\label{entropy_estimator_marginal}
   \phantom{,}\hat{H}_r(X) = - \langle \psi(n_x + 1) \rangle + \psi(N) + d_X \langle \ln \tilde{\varepsilon} \rangle,
\end{equation}
\begin{equation}\label{entropy_estimator_joint}
   \phantom{.}\hat{H}_r(X;Y) = - \psi(k) + \psi(N) + (d_X + d_Y) \langle \ln \tilde{\varepsilon} \rangle.
\end{equation}

NMI is then calculated by substituting these marginal and joint entropy estimates into Equations \eqref{mi_via_entropy_definition} and \eqref{nmi_equation}.

\section{Proposed transformation}
\label{sec:proposed}
We begin by taking the logarithm of the normalization factor---denominator in \eqref{scaling_invariant_radius_equation}---and applying the product and power rules of logarithms, as described in \cite[p. 37]{finney2001}:
\begin{equation}\label{logarithm_of_normalization_factor}
   \phantom{.}\ln V = \frac{1}{d_X + d_Y} \left[ \ln \dfrac{1}{N} + \ln \sum_{i=1}^{N} {\varepsilon_i^{d_X + d_Y}} \right].
\end{equation}

Next, the largest k-NN radius $\varepsilon_{\text{max}}$ is identified:
\begin{equation}\label{largest_radius}
   \phantom{.}\varepsilon_{\text{max}} = \max_{1 \leq i \leq N}{\varepsilon_i}.
\end{equation}

Using the \textit{log-sum-exp formula} from \cite[p. 844]{press2007} and applying the quotient rule of logarithms from \cite[p. 37]{finney2001} equation \eqref{logarithm_of_normalization_factor} is refactored by factoring out the maximum value $\varepsilon_{\text{max}}$ from \eqref{largest_radius}:
\begin{equation}\label{logarithm_of_normalization_factor_after_refactor}
   \phantom{.}\ln V = \ln \varepsilon_{\text{max}} + \tfrac{1}{d_X + d_Y} \ln\left(\frac{\sum_{i=1}^{N}{\left( \frac{\varepsilon_i}{\varepsilon_{\text{max}}} \right)^{d_X + d_Y}}}{N}\right).
\end{equation}

Substituting the result from \eqref{logarithm_of_normalization_factor_after_refactor} back into the original equation \eqref{scaling_invariant_radius_equation}---see Equation~\eqref{scaling_invariant_radius_equation_proposed}---avoids numerical overflow during the calculation of the scaling-invariant radii, as all the terms $\sfrac{\varepsilon_i}{\varepsilon_{\text{max}}}$ are guaranteed to be less than or equal to one.

\begin{equation}\label{scaling_invariant_radius_equation_proposed}
   \tilde{\varepsilon} = \varepsilon / {\rm e}^{\ln V}
 \end{equation}

Additionally, for very large values of $d_X + d_Y$, the scaling-invariant radii calculation is dominated by $\varepsilon_{\text{max}}$---as illustrated in Figure~\ref{fig:asymptotic_comparison}---since the second term in Equation~\eqref{logarithm_of_normalization_factor_after_refactor} becomes negligible. This happens because $\sfrac{\varepsilon_i}{\varepsilon_{\text{max}}}$ for $\varepsilon_i < \varepsilon_{\text{max}}$ decay exponentially to zero as $d_X + d_Y \to \infty$, effectively eliminating their contribution to the sum. However, the terms where $\varepsilon_i = \varepsilon_{\text{max}}$ remain equal to $1$ and continue to contribute to the sum, ensuring it stays finite and preventing the logarithm from tending to minus infinity. Moreover, the factor \(\frac{1}{d_X + d_Y} \to 0\) as $d_X + d_Y \to \infty$, which further diminishes the impact of the second term. As a result, Equation~\eqref{scaling_invariant_radius_equation_proposed} simplifies to $\tilde{\varepsilon} = \varepsilon / \varepsilon_{\text{max}}$ for large $d_X + d_Y$.

\section{Numerical Experiments}
\label{sec:experiments}
We evaluate the proposed transformation, which stabilizes scaling-invariant k-NN radius calculations while preserving precision, by comparing it to the baseline estimator that computes the radius directly, without transformation. Experiments cover varying correlation strengths, degrees of freedom, and dimensionalities. Results are benchmarked against theoretical NMI for standard Gaussian and heavy-tailed Student’s $t$ data, and against the Mutual Information Neural Estimator (MINE)~\cite{belghazi2018}, a widely used alternative based on a fundamentally different estimation paradigm. MINE outputs are normalized using Equation~\eqref{nmi_equation}, with marginal entropies estimated via Kernel Density Estimation (KDE)~\cite{rosenblatt1956,parzen1962} using Scott’s Rule~\cite{scott1992multivariate}, which provides a standard bias-variance tradeoff.

All settings use 10{,}000 samples and $K = 5$ neighbors for KSG-based estimators (baseline and proposed), following common practice. Each experiment is repeated 10 times; we report the means and standard deviations across runs.

\subsection{Synthetic Data and Ground-Truths}
\label{ssec:experimentalsetup}
We generate data from two families of distributions: multivariate Gaussian and Student's $t$. In both cases, we construct random vectors $(\mathbf{x}_i, \mathbf{y}_i) \in \mathbb{R}^d \times \mathbb{R}^d$ with component-wise correlation $\rho$ and zero mean. For the Gaussian case \cite{kullback1959information,cover2005}, the mutual information is given by \cite{belghazi2018}
\begin{equation}\label{mi_for_generated_data}
    \phantom{,}I_{\text{G}}(X; Y) = -\frac{d}{2} \ln(1 - \rho^2),
\end{equation}
and the marginal entropies simplify to \cite{cover2005}
\begin{equation}\label{entropy_for_generated_data_special_case}
    \phantom{,}H_{\text{G}}(X) = H_{\text{G}}(Y) = \frac{d}{2} \ln(2 \pi \mathrm{e}),
\end{equation}
yielding the normalized mutual information
\begin{equation}\label{true_nmi_for_generated_data}
    \phantom{.}I_{\text{NMI,G}}(X; Y) = -\frac{\ln(1 - \rho^2)}{\ln(2 \pi \mathrm{e})}.
\end{equation}
As $\rho$ approaches $1$, $\ln(1 - \rho^2)$ tends to $-\infty$, leading the Equation~\eqref{true_nmi_for_generated_data} to diverge. To ensure stability, we cap the NMI at $1.0$, which corresponds to the case of perfect correlation.

For the Student's $t$-distributed case \cite{mardia1979multivariate,kotz2004multivariate}, we generate $(X; Y)$ by first sampling latent Gaussian variables $(\tilde{X}, \tilde{Y}) \sim \mathcal{N}(0, \Omega)$ and an independent scalar $U \sim \chi^2_\nu$, then scaling via 
\[
  X = \tilde{X}\sqrt{\tfrac{\nu}{U}}, 
  \quad
  Y = \tilde{Y}\sqrt{\tfrac{\nu}{U}}.
\]
This introduces heavy-tailed dependencies controlled by the degrees of freedom $\nu$. The true mutual information is \cite{arellanovalle2013}
\begin{equation}\label{mi_for_generated_data_student}
    I_{\text{T}}(X; Y) = I(\tilde{X}; \tilde{Y}) + c(\nu, d),
\end{equation}
where
\begin{align}
  c(\nu, d) &= f(\nu) + f(\nu + 2d) - 2\,f(\nu + d), \label{c_definition} \\
  f(x)      &= \ln\Gamma\bigl(\tfrac{x}{2}\bigr) \;-\; \tfrac{x}{2}\,\psi\bigl(\tfrac{x}{2}\bigr). \label{f_definition}
\end{align}
Here, $\Gamma(\cdot)$ is the gamma function and $\psi(\cdot)$ is the digamma function. Notably, even when $\Omega = \mathbf{I}_{d}$, this $c(\nu,d)$ term makes $I_{\text{T}}(X;Y)\neq 0$ due to information encoded in the scaling variable $U$. The marginal entropies are given by \cite{arellanovalle2013}
\begin{equation}
    H_{\text{T}}(X) = H_{\text{T}}(Y) = \frac{d}{2} \ln(\nu \pi) + f(\nu) - f(\nu + d),
\end{equation}
yielding the normalized mutual information
\begin{equation}\label{true_nmi_for_generated_data_student}
    I_{\text{NMI,T}}(X; Y) = \frac{I(\tilde{X}; \tilde{Y}) + c(\nu, d)}{\frac{d}{2} \ln(\nu \pi) + f(\nu) - f(\nu + d)}.
\end{equation}

\subsection{Results}
\label{ssec:results}
\begin{figure*}
   \centering
   \includegraphics[width=\textwidth]{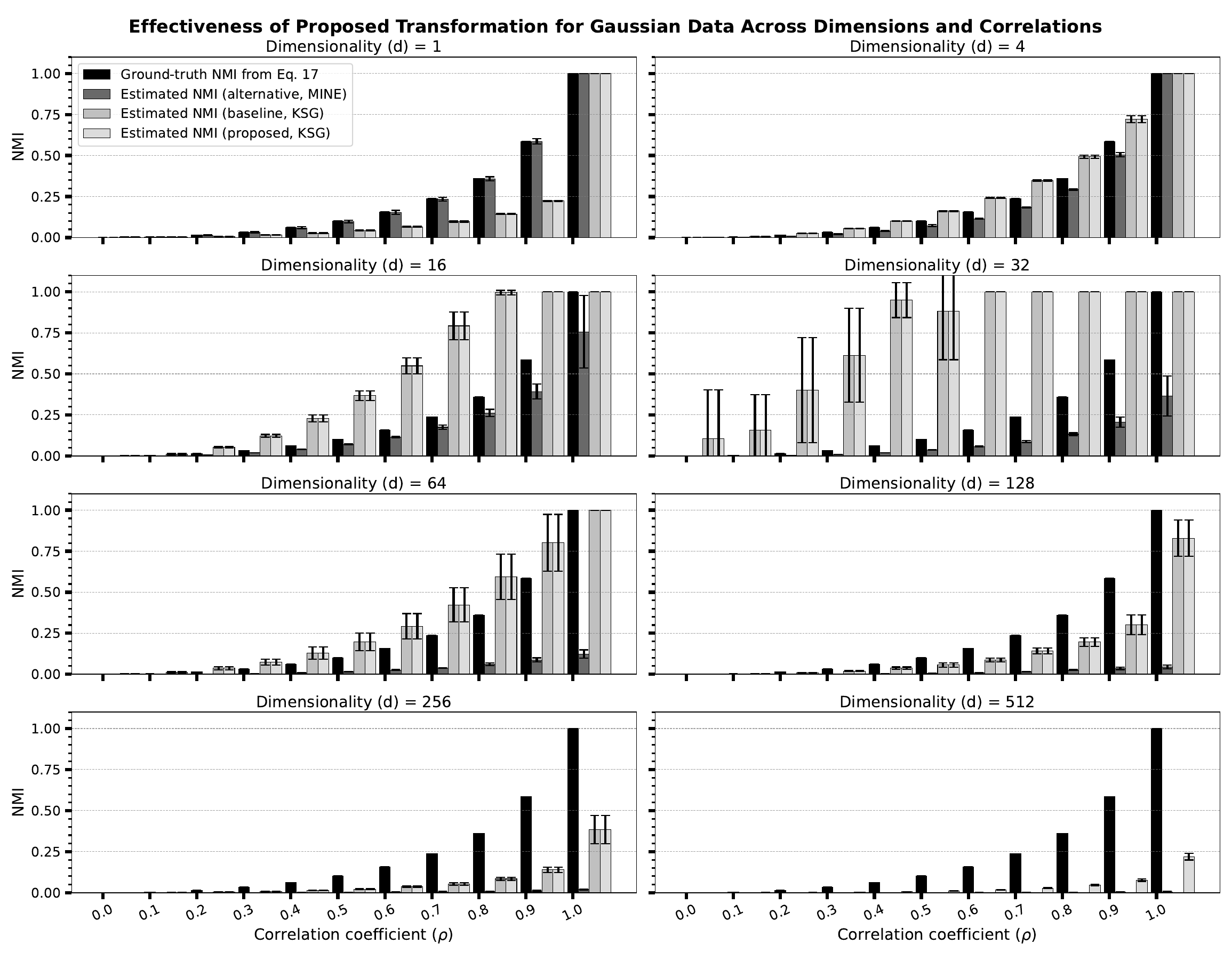}
   \caption{Estimated NMI for multivariate Gaussian data with joint covariance \cite{belghazi2018}, across dimensions $d = 1$ to $512$, plotted against correlation coefficient $\rho$ (ranging from $0$ to $1$ in steps of $0.1$). Each subplot corresponds to a dimensionality and shows the true NMI, along with estimates from the baseline method, the proposed method, and an alternative method. Bar height indicates the mean over 10 repetitions; black line segments denote standard deviation.}
   \label{fig:experimental_results_nmi_vs_rho}
\end{figure*}

Figure~\ref{fig:experimental_results_nmi_vs_rho} shows that for multivariate Gaussian data, the KSG-based estimators (baseline and proposed) tend to undershoot in low dimensions (1--2), overshoot in moderate ones (4--32), and again undershoot in high dimensions (64--512). The baseline estimator fails beyond 256 dimensions due to numerical overflow, while the proposed transformation remains stable throughout. Performance is more sensitive to dimensionality $d$ than to correlation $\rho$. Variance increases notably around 32 dimensions. The proposed and baseline estimators yield identical means and variances in all dimensions where the baseline remains numerically stable.

The MINE-based estimator (alternative) performs comparably or better than the KSG-based methods in low to moderate dimensions but fails to detect NMI in high-dimensional Gaussian data. In contrast, the baseline and proposed methods continue to recover some information, with the proposed method remaining functional even where the baseline fails.

\begin{figure*}
   \centering
   \includegraphics[width=\textwidth]{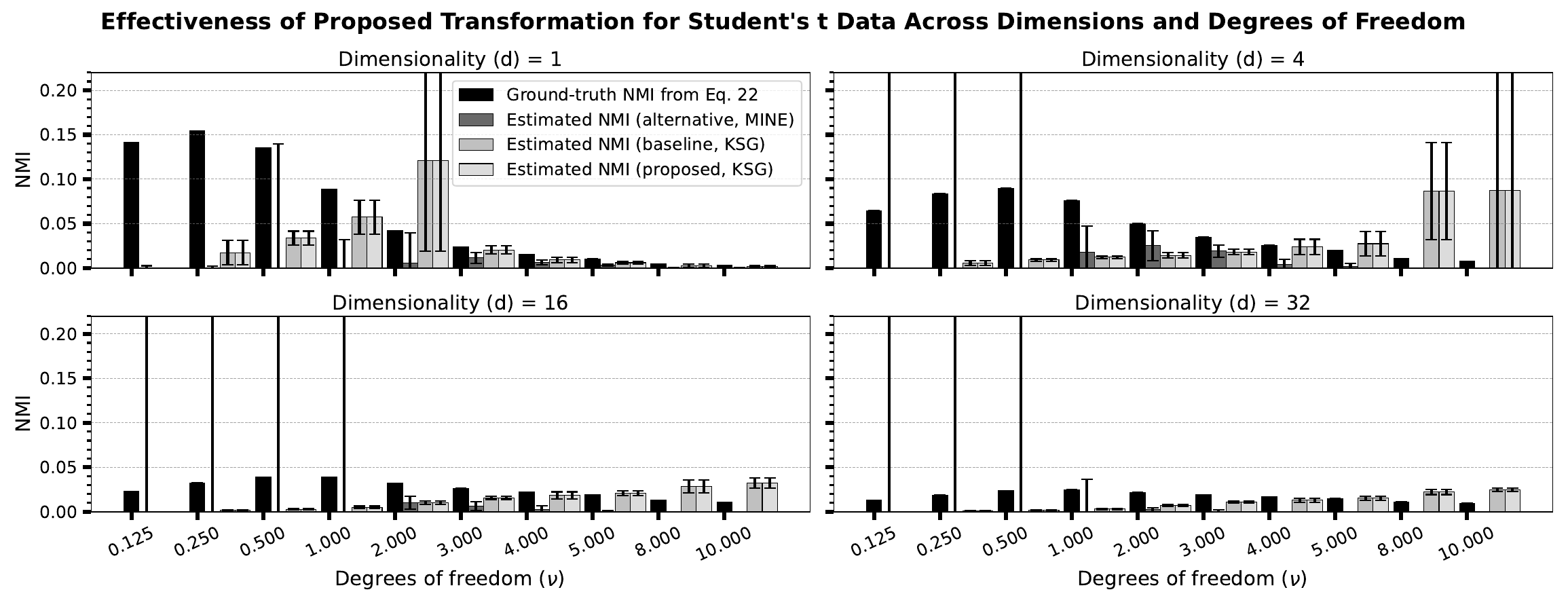}
   \caption{Estimated NMI for multivariate Student's $t$ data with identity dispersion matrix $\Omega$ \cite{czyz2023}, across dimensions $d = 1$ to $32$, plotted against degrees of freedom $\nu$ (ranging from $0.125$ to $10$). Each subplot corresponds to a dimensionality and shows the true NMI, along with estimates from the baseline method, the proposed method, and an alternative method. Bar height indicates the mean over 10 repetitions; black line segments denote standard deviation.}
   \label{fig:experimental_results_nmi_vs_dof}
\end{figure*}

Figure~\ref{fig:experimental_results_nmi_vs_dof} shows that under multivariate Student’s $t$-distributed data, all methods degrade when the tails are heavy (i.e., $\nu \leq 1$). The alternative estimator performs worst in this setting, frequently producing negative values and exhibiting extremely high variance. The baseline and proposed KSG-based estimators again behave identically in this regime.

These findings align with prior work~\cite{czyz2023}, underscoring the challenges of NMI estimation in high-dimensional and heavy-tailed settings. The proposed transformation improves numerical stability where the baseline fails and the alternative estimator breaks down.

\section{Conclusion}
\label{ssec:conclusion}
Mutual information is a fundamental measure for quantifying statistical dependencies between variables. In this work, we addressed a key limitation in Normalized Mutual Information (NMI) estimation using the Kraskov-Stögbauer-Grassberger (KSG)-based estimator---specifically, numerical overflow in high-dimensional settings caused by instability in the scaling-invariant k-nearest neighbors radius computation.

We proposed a simple logarithmic transformation that resolves this issue without introducing bias or computational overhead. Empirical results on both standard and heavy-tailed data confirm that the method remains stable up to 512 dimensions and reduces estimator variance by preventing numerical failure. We also evaluated a Mutual Information Neural Estimator (MINE)-based alternative, which showed mixed performance---competitive in low to moderate dimensions but unreliable in high-dimensional or heavy-tailed settings.

In summary, our transformation improves the numerical robustness of KSG-based NMI estimation in high dimensions. While it does not address inherent statistical limitations---such as the curse of dimensionality and metric concentration---it extends the estimator’s practical utility by enabling numerically stable computation in hundreds of dimensions, offering a more reliable tool for high-dimensional data analysis.

\clearpage
\input{appendix_one_page}

\clearpage
\bibliographystyle{IEEEtran}
\bibliography{spl_article}

\end{document}

%% file: appendix_one_page.tex
\appendices
\section{Rigorous Mathematical Justification of the Dominance of $\varepsilon_{\max}$ in High Dimensions}

This appendix provides a theoretical justification for why the largest k-NN radius ($\varepsilon_{\max}$)dominates the normalization factor in scaling-invariant k-NN radius calculations as $D = d_X + d_Y \to \infty$. Standard logarithmic identities (see, e.g., \cite[pp.~36–37]{finney2001}) are used to simplify expressions throughout.

\subsection{Restating and Manipulating the Normalization Factor $V$}
We start with the normalization factor $V$ from \textbf{Equation~\eqref{scaling_invariant_radius_equation} of the main manuscript} by taking the natural logarithm of both sides and applying the \textbf{power rule of logarithms}:
\begin{equation}
    \ln V = \frac{1}{d_X + d_Y} \ln \left(\frac{1}{N} \sum_{i=1}^{N} \varepsilon_i^{d_X + d_Y}\right).
\end{equation}

Expanding the logarithm on the right-hand side using the \textbf{product rule of logarithms}, and denoting $d_X + d_Y$ as $D$:
\begin{equation}
    \ln V = \frac{1}{D} \left[\ln \frac{1}{N} + \ln \sum_{i=1}^{N} \varepsilon_i^{D} \right].
    \label{eq:ln_v}
\end{equation}

\subsection{Factoring Out the Dominant Term ($\varepsilon_{\max}$)}
Define the largest k-NN radius:
\begin{equation}
    \varepsilon_{\max} = \max_{1 \leq i \leq N} \varepsilon_i.
    \label{eq:largest_epsilon}
\end{equation}

Drawing on the \textbf{log-sum-exp trick}---a numerically stable method for computing the logarithm of a sum of exponentials \cite[p.~844]{press2007}; see also standard machine learning texts \cite{murphy2022, goodfellow2016deep, bishop2006pattern}---we can rewrite the summation term as:
\begin{align}
    \sum_{i=1}^{N} \varepsilon_i^{D}
    &= \sum_{i=1}^{N} \left[ \frac{\varepsilon_{\max}^{D}}{\varepsilon_{\max}^{D}} \varepsilon_i^{D} \right] \\
    &= \sum_{i=1}^{N} \left[ \varepsilon_{\max}^{D} \frac{\varepsilon_i^{D}}{\varepsilon_{\max}^{D}} \right] \\
    &= \varepsilon_{\max}^{D} \sum_{i=1}^{N} \left(\frac{\varepsilon_i}{\varepsilon_{\max}}\right)^{D}.
\end{align}

Taking the natural logarithm of both sides and applying the \textbf{product rule of logarithms}:
\begin{equation}
    \ln \sum_{i=1}^{N} \varepsilon_i^{D} = \ln \varepsilon_{\max}^{D} + \ln \sum_{i=1}^{N} \left(\frac{\varepsilon_i}{\varepsilon_{\max}}\right)^{D}.
    \label{eq:refactored_summation}
\end{equation}

Substituting Equation~\eqref{eq:refactored_summation} back into Equation~\eqref{eq:ln_v}:
\begin{equation}
    \ln V = \frac{1}{D} \left[\ln \frac{1}{N} + \ln \varepsilon_{\max}^{D} + \ln \sum_{i=1}^{N} \left(\frac{\varepsilon_i}{\varepsilon_{\max}}\right)^{D}\right].
\end{equation}

Moving the term $\ln \varepsilon_{\max}^{D}$ outside the parentheses, applying the \textbf{power rule of logarithms}, and simplifying further:
\begin{align}
    \ln V &= \ln \varepsilon_{\max} + \frac{1}{D} \left[\ln \frac{1}{N} + \ln \sum_{i=1}^{N} \left(\frac{\varepsilon_i}{\varepsilon_{\max}}\right)^{D}\right].
    \label{eq:ln_v_refactored_final}
\end{align}

\subsection{Analyzing the Second Term in High Dimensions}
We focus on the second term in Equation~\eqref{eq:ln_v_refactored_final}:
\begin{equation}
    T = \frac{1}{D} \left[\ln \frac{1}{N} + \ln \sum_{i=1}^{N} \left(\frac{\varepsilon_i}{\varepsilon_{\max}}\right)^{D}\right]
    \label{eq:ln_v_second_term}
\end{equation}

\subsubsection{Behavior of Individual Terms in Summation as $D \to \infty$}
\begin{enumerate}
    \item If $\varepsilon_i < \varepsilon_{\max}$, then $\frac{\varepsilon_i}{\varepsilon_{\max}} < 1$. Raising a value less than 1 to a large power results in exponential decay:
    \begin{equation}
        \left(\frac{\varepsilon_i}{\varepsilon_{\max}}\right)^{D} \to 0 \quad \text{as} \quad \varepsilon_i < \varepsilon_{\max} \text{ and } D \to \infty.
    \end{equation}
    \item If $\varepsilon_i = \varepsilon_{\max}$, then $\frac{\varepsilon_i}{\varepsilon_{\max}} = 1$. Raising $1$ to any power equals to $1$:
        \begin{equation}
        \left(\frac{\varepsilon_i}{\varepsilon_{\max}}\right)^{D} \to 1 \quad \text{as} \quad \varepsilon_i = \varepsilon_{\max} \text{ and } D \to \infty.
    \end{equation}
\end{enumerate}

\subsubsection{Overall Contribution to the Summation as $D \to \infty$}
Let $n$ denote the number of terms where $\varepsilon_i = \varepsilon_{\max}$. Then:
\begin{equation}
    \sum_{i=1}^{N} \left(\frac{\varepsilon_i}{\varepsilon_{\max}}\right)^{D} \to n \quad \text{as} \quad D \to \infty,
\end{equation}
and the second term in Equation~\eqref{eq:ln_v_refactored_final} simplifies to $T \approx \frac{1}{D} \left[\ln \frac{1}{N} + \ln n\right]$. As $D \to \infty$, $\frac{1}{D} \to 0$, and thus
\begin{equation}
    T \to 0 \cdot \left[\ln \frac{1}{N} + \ln n\right] = 0,
    \label{eq:ln_v_second_term_limit}
\end{equation}
since $\ln \sfrac{1}{N} + \ln n$ remains finite.

\subsection{Approximation of Normalization Factor $V$ as $D \to \infty$}
Substituting limit from Equation~\eqref{eq:ln_v_second_term_limit} into Equation~\eqref{eq:ln_v_refactored_final}:
\begin{equation}
    \ln V = \ln \varepsilon_{\max} + T \to \ln \varepsilon_{\max} \quad \text{as} \quad D \to \infty.
\end{equation}

Exponentiating both sides and using the \textbf{inverse property of the logarithm and exponential}, we obtain:
\begin{align}
    V        &\to \varepsilon_{\max}, \quad \text{as} \quad D \to \infty.
    \label{eq:v_limit}
\end{align}

Substituting Equation~\eqref{eq:v_limit} into the equation of scaling-invariant radius:
\begin{equation}
    \tilde{\varepsilon} = \frac{\varepsilon}{V} \to \frac{\varepsilon}{\varepsilon_{\max}} \quad \text{as} \quad D \to \infty.
    \label{eq:scaling_invariant_radius_limit}
\end{equation}

\subsection{Interpretation and Conclusion}
\begin{enumerate}
    \item \textbf{Dominance of $\varepsilon_{\max}$:} As $D \to \infty$, all $\varepsilon_i < \varepsilon_{\max}$ become negligible, leaving only the largest k-NN radius, $\varepsilon_{\max}$, to influence the normalization factor $V$.    
    \item \textbf{Simplified Radius:} In this limit, as shown in Equation~\eqref{eq:scaling_invariant_radius_limit}, the scaling-invariant radius approximates as $\tilde{\varepsilon} \approx \varepsilon / \varepsilon_{\max}$.
    \item \textbf{Numerical Stability:} The logarithmic transformation ensures all terms remain bounded, improving numerical stability and avoiding overflow or underflow in high dimensions, i.e., when $D =  d_X + d_Y$ goes to infinity.
\end{enumerate}

\subsection{Final Remarks}
This theoretical derivation rigorously explains why the largest k-NN radius ($\varepsilon_{\max}$), as defined in Equation~\eqref{eq:largest_epsilon}, dominates the normalization factor $V$ in high-dimensional spaces, i.e., when $D =  d_X + d_Y$ goes to infinity.